\documentclass[10pt,a4paper,onecolumn]{article}
\usepackage{marginnote}
\usepackage{graphicx}
\usepackage{xcolor}
\usepackage{authblk,etoolbox}
\usepackage{titlesec}
\usepackage{calc}
\usepackage{tikz}
\usepackage{hyperref}
\hypersetup{colorlinks,breaklinks=true,
            urlcolor=[rgb]{0.0, 0.5, 1.0},
            linkcolor=[rgb]{0.0, 0.5, 1.0}}
\usepackage{caption}
\usepackage{tcolorbox}
\usepackage{amssymb,amsmath}
\usepackage{ifxetex,ifluatex}
\usepackage{seqsplit}
\usepackage{xstring}

\usepackage{float}
\let\origfigure\figure
\let\endorigfigure\endfigure
\renewenvironment{figure}[1][2] {
    \expandafter\origfigure\expandafter[H]
} {
    \endorigfigure
}

\usepackage{fixltx2e} 
\usepackage[
  backend=biber,
]{biblatex}
\bibliography{paper.bib}

\let\textttOrig=\texttt
\def\texttt#1{\expandafter\textttOrig{\seqsplit{#1}}}
\renewcommand{\seqinsert}{\ifmmode
  \allowbreak
  \else\penalty6000\hspace{0pt plus 0.02em}\fi}

\makeatletter
\let\href@Orig=\href
\def\href@Urllike#1#2{\href@Orig{#1}{\begingroup
    \def\Url@String{#2}\Url@FormatString
    \endgroup}}
\def\href@Notdoi#1#2{\def\tempa{#1}\def\tempb{#2}%
  \ifx\tempa\tempb\relax\href@Urllike{#1}{#2}\else
  \href@Orig{#1}{#2}\fi}
\def\href#1#2{%
  \IfBeginWith{#1}{https://doi.org}%
  {\href@Urllike{#1}{#2}}{\href@Notdoi{#1}{#2}}}
\makeatother

\newlength{\cslhangindent}
\newlength{\csllabelwidth}
\newenvironment{CSLReferences}[3] 
 {
  \setlength{\parindent}{0pt}
  \ifodd #1 \everypar{\setlength{\hangindent}{\cslhangindent}}\ignorespaces\fi
  \ifnum #2 > 0
  \setlength{\parskip}{#2\baselineskip}
  \fi
 }%
 {}
\usepackage{calc}

\usepackage[top=3.5cm, bottom=3cm, right=1.5cm, left=1.0cm,
            headheight=2.2cm, reversemp, includemp, marginparwidth=4.5cm]{geometry}

\titleformat{\section}
  {\normalfont\sffamily\Large\bfseries}
  {}{0pt}{}
\titleformat{\subsection}
  {\normalfont\sffamily\large\bfseries}
  {}{0pt}{}
\titleformat{\subsubsection}
  {\normalfont\sffamily\bfseries}
  {}{0pt}{}
\titleformat*{\paragraph}
  {\sffamily\normalsize}

\usepackage{fancyhdr}
\makeatletter
\let\ps@plain\ps@fancy
\fancyheadoffset[L]{4.5cm}
\fancyfootoffset[L]{4.5cm}

\definecolor{linky}{rgb}{0.0, 0.5, 1.0}

\newtcolorbox{repobox}
   {colback=red, colframe=red!75!black,
     boxrule=0.5pt, arc=2pt, left=6pt, right=6pt, top=3pt, bottom=3pt}

\newcommand{\ExternalLink}{%
   \tikz[x=1.2ex, y=1.2ex, baseline=-0.05ex]{%
       \begin{scope}[x=1ex, y=1ex]
           \clip (-0.1,-0.1)
               --++ (-0, 1.2)
               --++ (0.6, 0)
               --++ (0, -0.6)
               --++ (0.6, 0)
               --++ (0, -1);
           \path[draw,
               line width = 0.5,
               rounded corners=0.5]
               (0,0) rectangle (1,1);
       \end{scope}
       \path[draw, line width = 0.5] (0.5, 0.5)
           -- (1, 1);
       \path[draw, line width = 0.5] (0.6, 1)
           -- (1, 1) -- (1, 0.6);
       }
   }

\patchcmd{\@maketitle}{center}{flushleft}{}{}
\patchcmd{\@maketitle}{center}{flushleft}{}{}
\patchcmd{\@maketitle}{\LARGE}{\LARGE\sffamily}{}{}
\def\maketitle{{%
  
  \AB@maketitle}}
\makeatletter
\renewcommand\AB@affilsepx{ \protect\Affilfont}
\renewcommand\AB@affilnote[1]{{\bfseries #1}\hspace{3pt}}
\renewcommand{\affil}[2][]%
   {\newaffiltrue\let\AB@blk@and\AB@pand
      \if\relax#1\relax\def\AB@note{\AB@thenote}\else\def\AB@note{#1}%
        \setcounter{Maxaffil}{0}\fi
        \begingroup
        \let\href=\href@Orig
        \let\texttt=\textttOrig
        \let\protect\@unexpandable@protect
        \def\thanks{\protect\thanks}\def\footnote{\protect\footnote}%
        \@temptokena=\expandafter{\AB@authors}%
        {\def\\{\protect\\\protect\Affilfont}\xdef\AB@temp{#2}}%
         \xdef\AB@authors{\the\@temptokena\AB@las\AB@au@str
         \protect\\[\affilsep]\protect\Affilfont\AB@temp}%
         \gdef\AB@las{}\gdef\AB@au@str{}%
        {\def\\{, \ignorespaces}\xdef\AB@temp{#2}}%
        \@temptokena=\expandafter{\AB@affillist}%
        \xdef\AB@affillist{\the\@temptokena \AB@affilsep
          \AB@affilnote{\AB@note}\protect\Affilfont\AB@temp}%
      \endgroup
       \let\AB@affilsep\AB@affilsepx
}
\makeatother

\renewcommand\Affilfont{\sffamily\small\mdseries}
\ifnum 0\ifxetex 1\fi\ifluatex 1\fi=0 
  \usepackage[T1]{fontenc}
  \usepackage[utf8]{inputenc}

\else 
  \ifxetex
    \usepackage{mathspec}
    \usepackage{fontspec}

  \else
    \usepackage{fontspec}
  \fi
  \defaultfontfeatures{Ligatures=TeX,Scale=MatchLowercase}

\fi
\IfFileExists{upquote.sty}{\usepackage{upquote}}{}
\IfFileExists{microtype.sty}{%
\usepackage{microtype}
\UseMicrotypeSet[protrusion]{basicmath} 
}{}

\usepackage{hyperref}
\hypersetup{unicode=true,
            pdftitle={pgmuvi: Quick and easy Gaussian Process Regression for multi-wavelength astronomical timeseries},
            pdfborder={0 0 0},
            breaklinks=true}
\let\addcontentslineOrig=\addcontentsline
\def\addcontentsline#1#2#3{\bgroup
  \let\texttt=\textttOrig\addcontentslineOrig{#1}{#2}{#3}\egroup}
\let\markbothOrig\markboth
\def\markboth#1#2{\bgroup
  \let\texttt=\textttOrig\markbothOrig{#1}{#2}\egroup}
\let\markrightOrig\markright
\def\markright#1{\bgroup
  \let\texttt=\textttOrig\markrightOrig{#1}\egroup}

\IfFileExists{parskip.sty}{%
\usepackage{parskip}
}{
\setlength{\parindent}{0pt}
\setlength{\parskip}{6pt plus 2pt minus 1pt}
}
\providecommand{\tightlist}{%
  \setlength{\itemsep}{0pt}\setlength{\parskip}{0pt}}
\ifx\paragraph\undefined\else
\let\oldparagraph\paragraph
\renewcommand{\paragraph}[1]{\oldparagraph{#1}\mbox{}}
\fi
\ifx\subparagraph\undefined\else
\let\oldsubparagraph\subparagraph
\renewcommand{\subparagraph}[1]{\oldsubparagraph{#1}\mbox{}}
\fi

\title{\texttt{pgmuvi}: Quick and easy Gaussian Process Regression for
multi-wavelength astronomical timeseries}

        \author[1, 2]{Peter Scicluna}
          \author[3, 4]{Stefan Waterval}
          \author[5]{Diego A. Vasquez-Torres}
          \author[5]{Sundar Srinivasan}
          \author[6]{Sara Jamal}
    
      \affil[1]{European Southern Observatory, Alonso de Córdova 3107,
Vitacura, Santiago, Chile}
      \affil[2]{Space Science Institute, 4750 Walnut Street, Suite 205,
Boulder, CO 80301, USA}
      \affil[3]{New York University Abu Dhabi, PO Box 129188, Abu Dhabi,
United Arab Emirates}
      \affil[4]{Center for Astro, Particle and Planetary Physics
(CAP\(^3\)), New York University Abu Dhabi, PO Box 129188, Abu Dhabi,
United Arab Emirates}
      \affil[5]{IRyA, Universidad Nacional Autónoma de México, Morelia,
Michoacán, México}
      \affil[6]{Max Planck Institute for Astronomy, Königstuhl 17, 69117
Heidelberg, Germany}
  \date{\vspace{-7ex}}

\begin{document}
\maketitle

\marginpar{

  \begin{flushleft}
  \sffamily\small

  {\bfseries DOI:} \href{https://doi.org/DOI unavailable}{\color{linky}{DOI unavailable}}

  \vspace{2mm}

  {\bfseries Software}
  \begin{itemize}
    \setlength\itemsep{0em}
    \item \href{N/A}{\color{linky}{Review}} \ExternalLink
    \item \href{NO_REPOSITORY}{\color{linky}{Repository}} \ExternalLink
    \item \href{DOI unavailable}{\color{linky}{Archive}} \ExternalLink
  \end{itemize}

  \vspace{2mm}

  \par\noindent\hrulefill\par

  \vspace{2mm}

  {\bfseries Editor:} \href{https://example.com}{Pending
Editor} \ExternalLink \\
  \vspace{1mm}
    {\bfseries Reviewers:}
  \begin{itemize}
  \setlength\itemsep{0em}
    \item \href{https://github.com/Pending Reviewers}{@Pending
Reviewers}
    \end{itemize}
    \vspace{2mm}

  {\bfseries Submitted:} N/A\\
  {\bfseries Published:} N/A

  \vspace{2mm}
  {\bfseries License}\\
  Authors of papers retain copyright and release the work under a Creative Commons Attribution 4.0 International License (\href{http://creativecommons.org/licenses/by/4.0/}{\color{linky}{CC BY 4.0}}).

  \end{flushleft}
}

\hypertarget{summary}{%
\section{Summary}\label{summary}}

Time-domain observations are increasingly important in astronomy, and
are often the only way to study certain objects. The volume of
time-series data is increasing dramatically as new surveys come online -
for example, the Vera Rubin Observatory will produce 15 terabytes of
data per night, and its Legacy Survey of Space and Time (LSST) is
expected to produce five-year lightcurves for \(>10^7\) sources, each
consisting of 5 photometric bands. Historically, astronomers have worked
with Fourier-based techniques such as the Lomb-Scargle periodogram or
information-theoretic approaches; however, in recent years Bayesian and
data-driven approaches such as Gaussian Process Regression (GPR) have
gained traction. However, the computational complexity and steep
learning curve of GPR has limited its adoption. \texttt{pgmuvi} makes
GPR of multi-band timeseries accessible to astronomers by building on
cutting-edge open-source machine-learning libraries, and hence
\texttt{pgmuvi} retains the speed and flexibility of GPR while being
easy to use. It provides easy access to GPU acceleration and Bayesian
inference of the hyperparameters (e.g.~the periods), and is able to
scale to large datasets.

\hypertarget{statement-of-need}{%
\section{Statement of need}\label{statement-of-need}}

Astronomical objects are in general not static, but vary in brightness
over time. This is especially true for objects that are variable by
nature, such as pulsating stars, or objects that are variable due to
their orbital motion, such as eclipsing binaries. The study of these
objects is called time-domain astronomy, and is a rapidly growing field.
A wide range of approaches to time-series analysis have been developed,
ranging from simple period-finding algorithms to more complex machine
learning techniques (e.g. Donoso-Oliva et al., 2023; Friedman, 1984;
Huijse et al., 2018; Palmer, 2009 and many more). Perhaps the most
popular in astronomy is the Lomb-Scargle periodogram (Lomb, 1976;
Scargle, 1982), which is a Fourier-based technique to find periodic
signals in unevenly sampled data. However, the handling of unevenly
sampled data is not the only challenge in time-series analysis.

A particular challenge in astronomy is handling heterogeneous,
multiwavelength data (VanderPlas \& Ivezić, 2015). Data must often be
combined from a wide variety of instruments, telescopes or surveys, and
so the systematics or noise properties of different datasets vary
widely. In addition, by combining multiple wavelengths, we gain a better
understanding of the physical processes driving the variability of the
object. For example, some variability mechanisms differ as a function of
wavelength only in amplitude (e.g.~eclipsing binaries), while others may
vary in phase (e.g.~pulsating stars) or even period (e.g.~multiperiodic
systems). Thus, it is important to combine data from multiple
wavelengths in a way that accounts for these differences.

Gaussian processes (GPs) have recently become a popular tool to handle
these challenges. GPs are a flexible way to forward-model arbitrary
signals, by assuming that the signal is drawn from a multivariate
Gaussian distribution. By constructing a covariance function describing
the covariance between any two points in the signal, we can model the
signal as a Gaussian process. By doing so, we are freed from any
assumptions about sampling, and can model the signal as a continuous
function. We are also able to model the noise in the data, and thus
account for heteroscedastic noise. We also gain the ability to directly
handle multiple wavelengths, by constructing a multi-dimensional
covariance function. Hence, Gaussian Process Regression (GPR) is a
machine learning technique that is able to model non-periodic signals in
unevenly sampled data, and is thus well suited for the analysis of
astronomical time-series data.

However, GPR is not without its challenges. The most popular covariance
functions are often not tailored to modelling complex signals, which
implies that users must construct their own covariance functions. GPs
are also computationally expensive, and thus approximations must be used
to scale to large datasets. Finally, many GP packages either have very
steep learning curves or only provide limited features, and thus are not
suitable for the average astronomer.

In this paper we present a new Python package, \texttt{pgmuvi}, to
perform GPR on multi-wavelength astronomical time-series data. The
package is designed to be easy to use, and provide a quick way to
perform GPR on multi-wavelength data by exploiting uncommon but powerful
kernels and appoximations. The package is designed to be flexible and
allow the user to customize the GPR model to their needs.
\texttt{pgmuvi} exploits multiple strategies to scale regression to
large datasets, making it suitable for the current era of large-scale
astronomical surveys.

A number of other software packages exist to perform GPR, some of which,
such as \texttt{celerite} (D. Foreman-Mackey, 2018; D. Foreman-Mackey et
al., 2017), \texttt{tinygp} (Daniel Foreman-Mackey, 2023) or
\texttt{george} (Ambikasaran et al., 2015) were developed within the
astronomical community with astronomical time-series in mind. However,
these each have their own limitations. \texttt{celerite} is extremely
fast, but is limited to one-dimensional inputs, and thus cannot handle
multiwavelength data, except under certain restrictive assumptions.
Furthermore, since it achieves its speed through a specific form of
kernel decomposition, it is not able to handle arbitrary covariance
functions. It is therefore restricted to a small number of kernels with
specific forms; while it is able to handle the most common astronomical
timeseries by combining these kernels, not all signals can be modelled
in this way. \texttt{tinygp} is a more general package, able to retain a
high degree of flexibility while still being fast thanks to a
\texttt{JAX}-based implementation, whic makes it feasible to implement
models in \texttt{tinygp} that are equivalent to those in
\texttt{pgmuvi}. However, \texttt{pgmuvi} is designed to provide an
easier learning curve by packaging GPs with data transforms and
inference routines. In essence, \texttt{tinygp} could in principle be
used by \texttt{pgmuvi} as a GP backend instead of GPyTorch. For a
summary of the state of the art of GPR in astronomy, see the recent
review by Aigrain \& Foreman-Mackey (2023).

\texttt{pgmuvi} is used in two ongoing projects by our group: one of the
authors' (DAVT) masters thesis and the paper resulting from this work
deals with the analysis of multiwavelength light curves for targets from
the Nearby Evolved Stars Survey (NESS; Scicluna et al. (2022),
\url{https://evolvedstars.space}). This work served as the first test of
the code and has analyzed thousands of light curves at optical and
infrared wavelengths for over seven hundred dusty stars within 3 kpc of
the Solar Neighborhood. The paper will be published in 2023
(Vasquez-Torres et al., in prep.). A different project related to dusty
variable stars in M33 has also used \texttt{pgmuvi} to estimate the
periods of these objects from infrared light curves. This work will be
published in 2023 (Srinivasan et al., in prep.).

\hypertarget{method-and-features}{%
\section{Method and Features}\label{method-and-features}}

\texttt{pgmuvi} builds on the popular GPyTorch library. GPyTorch
(Gardner et al., 2018) is a Gaussian process library for PyTorch (Paszke
et al., 2019), which is a popular machine learning library for Python.
By default, \texttt{pgmuvi} exploits the highly-flexible Spectral
Mixture kernel (Wilson \& Adams, 2013) in GPyTorch, able to model a wide
variety of signals. This kernel is particularly interesting for
astronomical time-series data, as it can effectively model
multi-periodic and quasi-periodic signals. The spectral mixture kernel
models the power spectrum of the covariance matrix as Gaussian mixture
model (GMM), making it highly flexible, easy to interpret and adaptable
to multi-dimensional data. This kernel is known for its ability to
extrapolate, and is thus well suited to cases where prediction is
important (for example, preparing astronomical observations of variable
stars). By modelling the power spectrum in this way, \texttt{pgmuvi}
effectively filters out noise in the periodogram, and thus is able to
find the dominant periods in noisy data more effectively than, for
example, the Lomb-Scargle periodogram.

However, the flexibility of this kernel comes at a cost; for more than
one component in the mixture, the solution space becomes highly
non-convex, and the optimization of the kernel hyperparameters becomes
difficult. \texttt{pgmuvi} addresses this by first exploiting the
Lomb-Scargle periodogram to find the dominant periods in the data, and
then using these periods as initial guesses for the means of the mixture
components.

Multiple options are available to accelerate inference depending on the
size of the dataset. For small datasets, the exact GPs can be used to
scale to datasets of up to \(\sim1000\) points. \texttt{pgmuvi} can
exploit the Structured Kernel Interpolation (SKI) approximation (Wilson
\& Nickisch, 2015) to scale to datasets of up to \(\sim10^5\) points.
Future work will include implementing approximations for even larger
datasets: \texttt{pgmuvi} can in principle exploit the Sparse
Variational GP (SVGP) or Variational Nearest Neighbour (VNN)
approximations (Hensman et al., 2013; Wu et al., 2022) to scale to
datasets of arbitrary size. \texttt{pgmuvi} can employ also GPU
computing for both exact and variational GPs.

For exact GPs and SKI, \texttt{pgmuvi} performs maximum a posteriori
(MAP) estimation of the hyperparameters, or full Bayesian inference. MAP
estimation can exploit any PyTorch optimizer, but by default it uses
ADAM (Kingma \& Ba, 2014). Bayesian inference uses the \texttt{pyro}
(Bingham et al., 2018) implementation of the No-U-Turn Sampler (NUTS)
(Hoffman et al., 2014), which is a Hamiltonian Monte Carlo (HMC)
sampler.

Finally, by allowing arbitrary GPyTorch likelihoods to be used,
\texttt{pgmuvi} can be extended to a wide range of problems. For
example, an instance of \texttt{gpytorch.likelihoods.StudentTLikelihood}
can be dropped in to turn \texttt{pgmuvi} into a T-Process regressor, or
missing data can be handled using
\texttt{gpytorch.likelihoods.GaussianLikelihoodWithMissingObs}.

To summarise, the key features of \texttt{pgmuvi} are:

\begin{itemize}
\tightlist
\item
  Highly flexible kernel, able to model a wide range of multiwavelength
  signals
\item
  Able to exploit multiple strategies to scale to large datasets
\item
  GPU acceleration for both exact and variational GPs
\item
  Fully Bayesian inference using HMC
\item
  Initialisation of kernel hyperparameters using Lomb-Scargle
  periodogram
\item
  Automated creation of diagnostic and summary plots
\item
  Automated reporting of kernel hyperparameters and their uncertainties,
  and summary of MCMC chains.
\end{itemize}

\hypertarget{acknowledgements}{%
\section{Acknowledgements}\label{acknowledgements}}

This project was developed in part at the 2022 Astro Hack Week, hosted
by the Max Planck Institute for Astronomy and Haus der Astronomie in
Heidelberg, Germany. This work was partially supported by the Max Planck
Institute for Astronomy, the European Space Agency, the Gordon and Betty
Moore Foundation, the Alfred P. Sloan foundation.

SS and DAVT acknowledge support from UNAM-PAPIIT Program IA104822.

\hypertarget{references}{%
\section*{References}\label{references}}
\addcontentsline{toc}{section}{References}

\hypertarget{refs}{}
\begin{CSLReferences}{1}{0}
\leavevmode\hypertarget{ref-arev_2023_gps}{}%
Aigrain, S., \& Foreman-Mackey, D. (2023). Gaussian process regression
for astronomical time series. \emph{Annual Review of Astronomy and
Astrophysics}, \emph{61}(1), null.
\url{https://doi.org/10.1146/annurev-astro-052920-103508}

\leavevmode\hypertarget{ref-ambikasaran2015george}{}%
Ambikasaran, S., Foreman-Mackey, D., Greengard, L., Hogg, D. W., \&
O'Neil, M. (2015). {Fast Direct Methods for Gaussian Processes}.
\emph{IEEE Transactions on Pattern Analysis and Machine Intelligence},
\emph{38}, 252. \url{https://doi.org/10.1109/TPAMI.2015.2448083}

\leavevmode\hypertarget{ref-bingham2018pyro}{}%
Bingham, E., Chen, J. P., Jankowiak, M., Obermeyer, F., Pradhan, N.,
Karaletsos, T., Singh, R., Szerlip, P., Horsfall, P., \& Goodman, N. D.
(2018). {Pyro: Deep Universal Probabilistic Programming}. \emph{Journal
of Machine Learning Research}.

\leavevmode\hypertarget{ref-Donoso-Oliva2023-transformer}{}%
Donoso-Oliva, C., Becker, I., Protopapas, P., Cabrera-Vives, G., Vishnu,
M., \& Vardhan, H. (2023). {ASTROMER. A transformer-based embedding for
the representation of light curves}. \emph{670}, A54.
\url{https://doi.org/10.1051/0004-6361/202243928}

\leavevmode\hypertarget{ref-celerite2}{}%
Foreman-Mackey, D. (2018). {Scalable Backpropagation for Gaussian
Processes using Celerite}. \emph{Research Notes of the American
Astronomical Society}, \emph{2}(1), 31.
\url{https://doi.org/10.3847/2515-5172/aaaf6c}

\leavevmode\hypertarget{ref-tinygp}{}%
Foreman-Mackey, Daniel. (2023). \emph{{dfm/tinygp: The tiniest of
Gaussian Process libraries}} (Version v0.2.4rc1) {[}Computer
software{]}. Zenodo. \url{https://doi.org/10.5281/zenodo.7646759}

\leavevmode\hypertarget{ref-celerite1}{}%
Foreman-Mackey, D., Agol, E., Ambikasaran, S., \& Angus, R. (2017).
{Fast and Scalable Gaussian Process Modeling with Applications to
Astronomical Time Series}. \emph{154}, 220.
\url{https://doi.org/10.3847/1538-3881/aa9332}

\leavevmode\hypertarget{ref-supersmoother}{}%
Friedman, J. H. (1984). {A variable span scatterplot smoother}.
\emph{\emph{Laboratory for Computational Statistics, Stanford University
Technical Report}}, \emph{5}.

\leavevmode\hypertarget{ref-gardner2018gpytorch}{}%
Gardner, J., Pleiss, G., Weinberger, K. Q., Bindel, D., \& Wilson, A. G.
(2018). Gpytorch: Blackbox matrix-matrix gaussian process inference with
gpu acceleration. \emph{Advances in Neural Information Processing
Systems}, \emph{31}.

\leavevmode\hypertarget{ref-hensman2013gaussian}{}%
Hensman, J., Fusi, N., \& Lawrence, N. D. (2013). Gaussian processes for
big data. \emph{arXiv Preprint arXiv:1309.6835}.

\leavevmode\hypertarget{ref-hoffman2014no}{}%
Hoffman, M. D., Gelman, A., \& others. (2014). The no-u-turn sampler:
Adaptively setting path lengths in hamiltonian monte carlo. \emph{J.
Mach. Learn. Res.}, \emph{15}(1), 1593--1623.

\leavevmode\hypertarget{ref-Huijse_2018}{}%
Huijse, P., Estévez, P. A., Förster, F., Daniel, S. F., Connolly, A. J.,
Protopapas, P., Carrasco, R., \& Príncipe, J. C. (2018). Robust period
estimation using mutual information for multiband light curves in the
synoptic survey era. \emph{The Astrophysical Journal Supplement Series},
\emph{236}(1), 12. \url{https://doi.org/10.3847/1538-4365/aab77c}

\leavevmode\hypertarget{ref-kingma2014adam}{}%
Kingma, D. P., \& Ba, J. (2014). Adam: A method for stochastic
optimization. \emph{arXiv Preprint arXiv:1412.6980}.

\leavevmode\hypertarget{ref-Lomb1976}{}%
Lomb, N. R. (1976). {Least-Squares Frequency Analysis of Unequally
Spaced Data}. \emph{39}(2), 447--462.
\url{https://doi.org/10.1007/BF00648343}

\leavevmode\hypertarget{ref-Palmer_2009}{}%
Palmer, D. M. (2009). A FAST CHI-SQUARED TECHNIQUE FOR PERIOD SEARCH OF
IRREGULARLY SAMPLED DATA. \emph{The Astrophysical Journal},
\emph{695}(1), 496. \url{https://doi.org/10.1088/0004-637X/695/1/496}

\leavevmode\hypertarget{ref-pytorch}{}%
Paszke, A., Gross, S., Massa, F., Lerer, A., Bradbury, J., Chanan, G.,
Killeen, T., Lin, Z., Gimelshein, N., Antiga, L., Desmaison, A., Köpf,
A., Yang, E., DeVito, Z., Raison, M., Tejani, A., Chilamkurthy, S.,
Steiner, B., Fang, L., \ldots{} Chintala, S. (2019). PyTorch: An
imperative style, high-performance deep learning library. In
\emph{Proceedings of the 33rd international conference on neural
information processing systems}. Curran Associates Inc.

\leavevmode\hypertarget{ref-Scargle1982}{}%
Scargle, J. D. (1982). {Studies in astronomical time series analysis.
II. Statistical aspects of spectral analysis of unevenly spaced data.}
\emph{263}, 835--853. \url{https://doi.org/10.1086/160554}

\leavevmode\hypertarget{ref-Scicluna2022}{}%
Scicluna, P., Kemper, F., McDonald, I., Srinivasan, S., Trejo, A.,
Wallström, S. H. J., Wouterloot, J. G. A., Cami, J., Greaves, J., He,
J., Hoai, D. T., Kim, H., Jones, O. C., Shinnaga, H., Clark, C. J. R.,
Dharmawardena, T., Holland, W., Imai, H., van Loon, J. T., \ldots{}
Zijlstra, A. A. (2022). {The Nearby Evolved Stars Survey II:
Constructing a volume-limited sample and first results from the James
Clerk Maxwell Telescope}. \emph{512}(1), 1091--1110.
\url{https://doi.org/10.1093/mnras/stab2860}

\leavevmode\hypertarget{ref-VanderPlas2015}{}%
VanderPlas, J. T., \& Ivezić, Ž. (2015). {Periodograms for Multiband
Astronomical Time Series}. \emph{812}(1), 18.
\url{https://doi.org/10.1088/0004-637X/812/1/18}

\leavevmode\hypertarget{ref-wilson:2013}{}%
Wilson, A., \& Adams, R. (2013). Gaussian process kernels for pattern
discovery and extrapolation. In S. Dasgupta \& D. McAllester (Eds.),
\emph{Proceedings of the 30th international conference on machine
learning} (Vol. 28, pp. 1067--1075). PMLR.
\url{https://proceedings.mlr.press/v28/wilson13.html}

\leavevmode\hypertarget{ref-wilson2015kernel}{}%
Wilson, A., \& Nickisch, H. (2015). Kernel interpolation for scalable
structured gaussian processes (KISS-GP). \emph{International Conference
on Machine Learning}, 1775--1784.

\leavevmode\hypertarget{ref-wu2022variational}{}%
Wu, L., Pleiss, G., \& Cunningham, J. P. (2022). Variational nearest
neighbor gaussian process. \emph{International Conference on Machine
Learning}, 24114--24130.

\end{CSLReferences}

\end{document}